\documentclass[numreferences]{kluwer}    

\usepackage{amssymb}
\usepackage{euscript}
\usepackage{graphicx}

\newtheorem{definition}{Definition}
\newtheorem{example}{Example}
\newtheorem{theorem}{Theorem}

\newtheorem{algorithm_}{Algorithm}

\begin{document}                                                                                   
\begin{article}
\begin{opening}         
\title{Modelling the Eulerian path problem using a string matching framework}
\author{Drago\c s N. \surname{Trinc\u a}\email{dnt04001@engr.uconn.edu}}  
\runningauthor{Drago\c s N. Trinc\u a}
\runningtitle{Modelling the Eulerian path problem using a string matching framework}
\institute{Department of Computer Science \& Engineering, University of Connecticut\\Storrs, CT 06269, USA}

\begin{abstract}
	The well-known Eulerian path problem can be solved in polynomial time
	(more exactly, there exists a linear time algorithm for this problem) \cite{clrs:1}.
	In this paper, we model the problem using a string matching framework, and then initiate an
	algorithmic study on a variant of this problem, called
	the \textsf{$(2,1)$--STRING--MATCH} problem (which is actually a generalization of the
	Eulerian path problem).
	Then, we present a polynomial-time algorithm for the \textsf{$(2,1)$--STRING--MATCH}
	problem, which is the most important result of this paper.
	Specifically, we get a lower bound of $\Omega(n)$, and an upper bound of $O(n^{2})$.
\end{abstract}

\end{opening}

\section{Introduction}
	The \textsf{$(2,1)$--STRING--MATCH} problem, as it will be formulated below,
	has been frequently encountered in many areas of computer science, especially as the well-known
	Eulerian path problem (which is actually a particular case of the
	\textsf{$(2,1)$--STRING--MATCH} problem). In this paper, we initiate
	an algorithmic study on this variant of the Eulerian path problem. As we shall see throughout the paper, it can
	be solved in polynomial-time using some basic graph theory concepts.
	Let us first fix some basic terminology.
	
\textit{Basic notions and notation.}
	The set of natural numbers is denoted by $\mathbb{N}$.
	A \textit{multiset} is a $2$-uple $(X,f)$, where $X$ is a set and $f:X\rightarrow\mathbb{N}$ is a function.
	A finite and nonempty set is called \textit{alphabet}.
	If $\Sigma$ is an alphabet, then $\Sigma^{n}$ denotes the set of all strings of length $n$ over $\Sigma$.
	For a string $x=\sigma_{1}\ldots{\sigma_{n}}\in\Sigma^{n}$, let $\textit{First}(x,i)$ denote
	the substring $\sigma_{1}\ldots{\sigma_{i}}$, and let $\textit{Last}(x,i)$ denote the substring
	$\sigma_{n-i+1}\ldots{\sigma_{n}}$.
	Let $U=(u_{1},\ldots,u_{k})$ be a $k$-uple. We denote by $U.i$ the $i$-th component of $U$, that is,
	$U.i=u_{i}$ for all $i\in{\{1,\ldots,k\}}$.
	The $0$-uple is denoted by $()$.
	If $q$ is an element or an uple, and $i\in{\{1,\ldots,k\}}$, then
	we define the uples $U\vartriangleleft{q}$ and $U\vartriangleright{i}$ by:
\begin{itemize}
\item $U\vartriangleleft{q}=(u_{1},\ldots,u_{k},q);$
\item $U\vartriangleright{i}=(u_{1},\ldots,u_{i-1},u_{i+1},\ldots,u_{k}).$
\end{itemize}

	If $s,t\in\mathbb{N}$ such that $1\leq{t}<s$, then the \textsf{$(s,t)$--STRING--MATCH} problem is stated
	as follows.
\begin{description}
\item[Given:] An alphabet $\Sigma$, and a $n$-uple $U$, $n\geq{2}$, such that $U.i\in\Sigma^{s}$
	for all $i\in\{1,\ldots,n\}$.
\item[Output:] Does there exist a permutation $p=(j_{1},\ldots,j_{n})$ of the set
	$\{1,\ldots,n\}$ such that $\textit{Last}(U.j_{i},t)=\textit{First}(U.j_{i+1},t)$ for all $i\in\{1,\ldots,n-1\}$?
\end{description}
\begin{example}
	Let $\Sigma=\{a,b,c\}$, and let $U=(ab,ac,cb,cc,ba)$ be a $5$-uple. Consider the
	permutation $p=(1,5,2,4,3)$ of the set $\{1,2,3,4,5\}$. One can verify that $U.1=U.5$,
	$U.5=U.2$, $U.2=U.4$, and $U.4=U.3$, that is, $U$ is a ``$\textit{YES}$'' instance of the
	\textsf{\textup{$(2,1)$--STRING--MATCH}} problem.
\end{example}
	Throughout the paper, we study only the \textsf{$(2,1)$--STRING--MATCH} problem, since it can
	be easily modelled using the graph theory. However, we consider that some of the results presented
	may apply to the generalized case as well.
\section{A Polynomial-time Algorithm for the \textsf{$(2,1)$--STRING--MATCH} Problem}
	This section is organized as follows. First, we recall some basic definitions
	and reformulate the problem stated above using the graph theory. Then, we present two naive
	(and superpolynomial-time) algorithms
	for solving the \textsf{$(2,1)$--STRING--MATCH} problem.
	Finally, we prove the main result of this paper,
	which gives a polynomial-time algorithm to our problem.
\begin{definition}
	A \textup{pseudodigraph} is a $3$-uple $G=(V,E,f)$, where $V$ is the set of \textup{vertices},
	$E\subseteq{V\times{V}}$ is the set of \textup{edges}, and $(E,f)$ is a multiset.
	If a pseudodigraph is specified simply by $G$, we denote by $V(G)$ (or $G.1$) its set of vertices,
	and by $E(G)$ (or $G.2$) its set of edges.
\end{definition}
	If $(M,g)$ is a multiset such that $M\subseteq{E}$ and $g(e)\leq{f(e)}$ for all $e\in{M}$,
	then the \textit{subpseudodigraph} of $G$ induced by the multiset $(M,g)$ is the $3$-uple
	$H=(V_{1},M,g)$, where $V_{1}=\{v \mid \exists$ $e\in{M}$ such that $v\in{\{e.1,e.2\}}\}$.
	
	A \textit{path} in a pseudodigraph is an uple $P=(e_{1},\ldots,e_{k})$ of edges such that
	$e_{i}.2=e_{i+1}.1$ for all $i\in\{1,\ldots,k-1\}$.
	A \textit{chain} is an uple $C=(e_{1},\ldots,e_{k})$ of edges such that
	$\{e_{i}.1,e_{i}.2\}\cap\{e_{i+1}.1,e_{i+1}.2\}\neq\emptyset$ for all $i\in\{1,\ldots,k-1\}$.
	A pseudodigraph $G$ is called
	\textit{connected} if any two of its vertices are linked by a chain in $G$.

	If $G=(V,E,f)$ is a pseudodigraph, then $F^{+}(v)=\sum_{(v,v')\in{E}}f(v,v')$, and
	$F^{-}(v)=\sum_{(v',v)\in{E}}f(v',v)$.
\begin{definition}
	Let $\Sigma$ be an alphabet, and $U$ a $n$-uple such that $U.i\in\Sigma^{2}$ for all
	$i\in\{1,\ldots,n\}$. The \textup{pseudodigraph associated to $U$}, denoted by $G(U)$,
	is a pseudodigraph $G(U)=(V,E,f)$ such that:
\begin{itemize}
\item $V=\{v \mid \exists$ $i$ such that $v\in\{\textit{First}(U.i,1),\textit{Last}(U.i,1)\}\}$,
\item $E=\{(x,y) \mid \exists$ $i\in\{1,\ldots,n\}$ such that $U.i=xy\}$,
\item $f(x,y)=|\{j \mid U.j=xy\}|$ for all $(x,y)\in{E}$.
\end{itemize}
	It is easy to verify that $\sum_{e\in{E}}f(e)=n$.
\end{definition}
\begin{example}
	Let $\Sigma=\{a,b,c,d,e,f\}$ be an alphabet, and consider the $9$-uple
\begin{center}
	$U=(ca,eb,ad,bf,dc,fe,ab,ab,ba)$
\end{center}
	whose components are strings of length two over $\Sigma$.
	Then, $G(U)$ can be illustrated as in the following figure.
\begin{figure}[h]
\setlength{\unitlength}{1pt}
\begin{picture}(335,80)
	\put(120,40){\circle{15}} 
	\put(215,40){\circle{15}} 
	\put(30,60){\circle{15}}  
	\put(30,20){\circle{15}}  
	\put(305,60){\circle{15}} 
	\put(305,20){\circle{15}} 

	\put(117,37){$a$}
	\put(212,37){$b$}
	\put(27,57){$c$}
	\put(27,17){$d$}
	\put(302,57){$e$}
	\put(302,17){$f$}

	\put(207.5,40){\vector(-1,0){80}}
	\put(127,43.1){\vector(1,0){81}}
	\put(127,36.9){\vector(1,0){81}}
	\put(30,27.5){\vector(0,1){25}}
	\put(305,27.5){\vector(0,1){25}}
	\put(113,36.9){\rotatebox{12}{\vector(-1,0){77.2}}}
	\put(222,36.9){\rotatebox{348}{\vector(1,0){77}}}
	
	\put(37.3,59){\rotatebox{-12}{\vector(1,0){77.3}}}
	\put(297.2,59){\rotatebox{12}{\vector(-1,0){77.1}}}
\end{picture}
\caption{$G(U)$: the pseudodigraph associated to $U$.}
\end{figure}
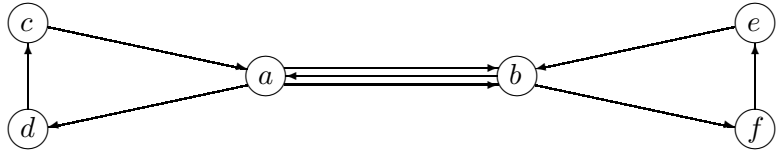
\end{example}
	In terms of graphs, the \textsf{$(2,1)$--STRING--MATCH} problem can be formulated as follows.
\begin{description}
\item[Given:] An alphabet $\Sigma$, a $n$-uple $U$ such that $U.i\in\Sigma^{2}$ for all
	$i\in\{1,\ldots,n\}$, and $G(U)=(V,E,f)$ the pseudodigraph
	associated to $U$.
\item[Output:] Does there exist a path $P=(e_{1},\ldots,e_{n})$ in $G(U)$ such that we have
	$f(e)=|\{j \mid e_{j}=e\}|$ for all $e\in{E}$?
\end{description}
\begin{algorithm_}
	\textup{(A naive algorithm)}
	Let $\Sigma$ be an alphabet, and $U$ a $n$-uple, $n\geq{2}$, such that $U.i\in\Sigma^{2}$ for all
	$i\in\{1,\ldots,n\}$. The simplest algorithm for the \textup{\textsf{$(2,1)$--STRING--MATCH}}
	problem is to determine whether there exists a permutation
	$p=(j_{1},\ldots,j_{n})$ of the set $\{1,\ldots,n\}$
	such that $U.j_{i}=U.j_{i+1}$ for all $i\in\{1,\ldots,n-1\}$. In the worst case, this algorithm
	runs in $O(n!)$, which is very slow.
\end{algorithm_}
\begin{algorithm_}
	Let us now describe a recursive algorithm, which can be obtained by improving the previous one.
	The idea is that we try to find a solution (that is, a permutation) progressively,
	by adding a new component to the current uple until we get an uple of length $n$.
	Note that the first call of the function must be \textup{StringMatch$1$($(),U,0,n$)}.
\textup{
\\\\
StringMatch$1$(uple $S$, uple $R$, integer $t$, integer $n$)
\\
$\texttt{ 1.}$ \textbf{begin}
\\
$\texttt{ 2.}$ \hspace{20pt}$res:=``\textit{NO}";$
\\
$\texttt{ 3.}$ \hspace{20pt}\textbf{if} $t=0$ \textbf{then}
\\
$\texttt{ 4.}$ \hspace{40pt}\textbf{for} $i=1$ \textbf{to} $n$ \textbf{do}
\\
$\texttt{ 5.}$ \hspace{40pt}\textbf{begin}
\\
$\texttt{ 6.}$ \hspace{60pt}$res:=$StringMatch$1$($(R.i),R\vartriangleright{i},1,n$)$;$
\\
$\texttt{ 7.}$ \hspace{60pt}\textbf{if} $res=``\textit{YES}"$ \textbf{then goto} $22.;$ 
\\
$\texttt{ 8.}$ \hspace{40pt}\textbf{end}
\\
$\texttt{ 9.}$ \hspace{20pt}\textbf{else}
\\
$\texttt{10.}$ \hspace{40pt}\textbf{for} $i=1$ \textbf{to} $n-t$ \textbf{do}
\\
$\texttt{11.}$ \hspace{60pt}\textbf{if} $\textit{Last}(S.t,1)=\textit{First}(R.i,1)$ \textbf{then}
\\
$\texttt{12.}$ \hspace{80pt}\textbf{if} $t=n-1$ \textbf{then}
\\
$\texttt{13.}$ \hspace{80pt}\textbf{begin}
\\
$\texttt{14.}$ \hspace{100pt}$res:=``\textit{YES}";$
\\
$\texttt{15.}$ \hspace{100pt}\textbf{goto} $22.;$
\\
$\texttt{16.}$ \hspace{80pt}\textbf{end}
\\
$\texttt{17.}$ \hspace{80pt}\textbf{else}
\\
$\texttt{18.}$ \hspace{80pt}\textbf{begin}
\\
$\texttt{19.}$ \hspace{100pt}$res:=$StringMatch$1$($S\vartriangleleft{R.i},R\vartriangleright{i},t+1,n$)$;$
\\
$\texttt{20.}$ \hspace{100pt}\textbf{if} $res=``\textit{YES}"$ \textbf{then goto} $22.;$
\\
$\texttt{21.}$ \hspace{80pt}\textbf{end}
\\
$\texttt{22.}$ \hspace{20pt}\textbf{return} $res;$
\\
$\texttt{23.}$ \textbf{end}
\\\\
}
	It is trivial to verify that in the worst case, the time complexity of this algorithm is superpolynomial.
	Specifically, we get an upper bound of $O(n!)$.
	We need a better algorithm. As we shall see, the \textsf{\textup{$(2,1)$--STRING--MATCH}} problem can
	be solved in polynomial time.
\end{algorithm_}
	Let us now prove the main result of this paper, which gives a polynomial-time algorithm for the
	\textsf{$(2,1)$--STRING--MATCH} problem.
\begin{theorem}
	Let $\Sigma$ be an alphabet, and let $U$ be a $n$-uple such that $U.i\in\Sigma^{2}$ for
	all $i\in\{1,\ldots,n\}$. Then $U$ is a ``\textit{YES}'' instance of the
	\textup{\textsf{$(2,1)$--STRING--MATCH}} problem
	if and only if $G(U)=(V,E,f)$ is connected and exactly one of the following two conditions holds:
\begin{enumerate}
\item[1.] $F^{+}(v)=F^{-}(v)$ for all $v\in{V}$; 
\item[2.] There exist $2$ vertices $v_{1},v_{2}\in{V}$, $v_{1}\neq{v_{2}}$, such that
	$F^{+}(v_{1})=F^{-}(v_{1})+1$, $F^{-}(v_{2})=F^{+}(v_{2})+1$, and $F^{+}(v)=F^{-}(v)$
	for all $v\in{V-\{v_{1},v_{2}\}}$.
\end{enumerate}
\begin{pf*}{Proof}
\\\\
\textup{Necessity. }
\\
	First, recall that we have $\sum_{e\in{E}}f(e)=n$.
	Let us assume that $U$ is a ``\textit{YES}'' instance of the \textup{\textsf{$(2,1)$--STRING--MATCH}} problem,
	that is, there exists a path $P=(e_{1},\ldots,e_{n})$ in $G(U)$ such that
	$f(e)=|\{j \mid e_{j}=e\}|$ for all $e\in{E}$.

	Given that $P$ is a path of length $n$, $\sum_{e\in{E}}f(e)=n$,
	and $f(e)=|\{j \mid e_{j}=e\}|$ for all $e\in{E}$,
	we get that $(E,f)=\{e_{1},\ldots,e_{n}\}$. This implies that the set $V$ cannot be divided into
	$2$ disjoint sets $X,Y$ such that $\{(x,y) \mid x\in{X}$ and $y\in{Y}\}\cap{E}=\emptyset$.
	Thus, we get that $G(U)$ is connected.

	Since $(E,f)=\{e_{1},\ldots,e_{n}\}$ and $e_{i}.2=e_{i+1}.1$ for all $i\in\{1,\ldots,n-1\}$,
	it follows that $F^{+}(v)=F^{-}(v)$ for all $v\in{V-\{e_{1}.1,e_{n}.2\}}$.
	If $e_{1}.1=e_{n}.2$ then we have $F^{+}(e_{1}.1)=F^{-}(e_{1}.1)$, and so we are
	in the first case. Otherwise, if $e_{1}.1\neq{e_{n}.2}$, we get that
	$F^{+}(e_{1}.1)=F^{-}(e_{1}.1)+1$ and $F^{-}(e_{n}.2)=F^{+}(e_{n}.2)+1$, that is, the second case.
\\\\
\textup{Sufficiency. }
\newline
	1. 
	Let us assume that $G(U)$ is connected and we are in the first case, that is,
	$F^{+}(v)=F^{-}(v)$ for all $v\in{V}$.
	Let $P_{1}=(e_{1},\ldots,e_{p})$ be a path of maximal length in $G(U)$ such that
	$p\leq{n}$ and $f(e)\geq{|\{j \mid e_{j}=e\}|}$ for all $e\in{E}$.
	If $p=n$ then $f(e)=|\{j \mid e_{j}=e\}|$ for all $e\in{E}$, and thus,
	we conclude that $U$ is a ``\textit{YES}'' instance of the \textup{\textsf{$(2,1)$--STRING--MATCH}} problem.
	
	Otherwise, if $p<n$, it follows that there exists an edge $(x,y)\in{E}$ such that
	$f(x,y)>|\{j \mid e_{j}=(x,y)\}|$. Since $F^{+}(v)=F^{-}(v)$ for all $v\in{V}$,
	$f(e)\geq{|\{j \mid e_{j}=e\}|}$ for all $e\in{E}$, and $P_{1}$ is a path of maximal length,
	we find that $e_{1}.1=e_{p}.2$ (otherwise, if $e_{1}.1\neq{e_{p}.2}$,
	it follows that there exists an edge $(e_{p}.2,z)\in{E}$ such that	
	$f(e_{p}.2,z)>|\{j \mid e_{j}=(e_{p}.2,z)\}|$, that is, $P_{1}$ is not a path of maximal length
	with the two properties specified above).

	Since $G(U)$ is connected, we conclude that there exists an edge $(x,y)\in{E}$ and $i\in\{1,\ldots,p\}$
	such that $f(x,y)>|\{j \mid e_{j}=(x,y)\}|$ and $x\in{\{e_{i}.1,e_{i}.2\}}$.
	Let $x=e_{i}.1$, and denote by $G_{1}=(V_{1},E_{1},f_{1})$ the subpseudodigraph of $G(U)$
	induced by the edges $e\in{E}$ with $f(e)>|\{j \mid e_{j}=e\}|$. Exactly, we have:
\begin{itemize}
\item $V_{1}=\{v \mid \exists$ $e$ such that $v\in\{e.1,e.2\}$ and $f(e)>|\{j \mid e_{j}=e\}|\}$,
\item $E_{1}=\{e \mid e\in{E}$ and $f(e)>|\{j \mid e_{j}=e\}|\}$,
\item $f_{1}(e)=f(e)-|\{j \mid e_{j}=e\}|$ for all $e\in{E_{1}}$.
\end{itemize}
	Using the fact that $F^{+}(v)=F^{-}(v)$ for all $v\in{V}$, $G(U)$ is connected, $P_{1}$ is
	a path, and $e_{1}.1=e_{p}.2$, we find that there exists a path $P_{2}=(z_{1},\ldots,z_{q})$
	in $G_{1}$ such that $z_{1}=(x,y)$, $z_{q}.2=x$, and $f_{1}(e)\geq{|\{j \mid z_{j}=e\}|}$
	for all $e\in{E_{1}}$.
	
	If we consider the path
	$P=(e_{1},\ldots,e_{i-1},z_{1},\ldots,z_{q},e_{i},\ldots,e_{p})$, we find that
	$f(e)\geq{|\{j \mid P.j=e\}|}$ for all $e\in{E}$, and $p+q\leq{n}$, that is, $P_{1}$ is not a path of maximal
	length in $G(U)$. Thus, assuming that $p<n$, we get a contradiction.
\\
	2. 
	Let us assume that $G(U)$ is connected and we are in the second case, that is,
	there exist $2$ vertices $v_{1},v_{2}\in{V}$, $v_{1}\neq{v_{2}}$, such that
	$F^{+}(v_{1})=F^{-}(v_{1})+1$, $F^{-}(v_{2})=F^{+}(v_{2})+1$,
	and $F^{+}(v)=F^{-}(v)$ for all $v\in{V-\{v_{1},v_{2}\}}$.

	Let $Z=U\vartriangleleft{v_{2}v_{1}}$ be a $(n+1)$-uple. Then, one can easily verify that
	$G(Z)$ is connected and $F^{+}(v)=F^{-}(v)$ for all $v\in{G(Z).1}$.
	According to the first case, $Z$ is a ``\textit{YES}'' instance of the
	\textup{\textsf{$(2,1)$--STRING--MATCH}} problem.
	
	If $G(Z)=(V_{Z},E_{Z},f_{Z})$, then let $P=(e_{1},\ldots,e_{n+1})$ be a path in $G(Z)$ such that
	$f_{Z}(e)=|\{j \mid P.j=e\}|$ for all $e\in{E_{Z}}$. Also, it is easy to verify
	that $e_{1}.1=e_{n+1}.2$. Let $i\in\{1,\ldots,n+1\}$ be such that $e_{i}=(v_{2},v_{1})$.
	Then, the $n$-uple $T=(e_{i+1},\ldots,e_{n+1},e_{1},\ldots,e_{i-1})$ is a path in $G(U)$
	and $f(e)=|\{j \mid T.j=e\}|$ for all $e\in{E}$. Thus, $U$ is
	a ``\textit{YES}'' instance of the \textup{\textsf{$(2,1)$--STRING--MATCH}} problem.
\qed
\end{pf*}
\end{theorem}
\begin{algorithm_}
	Let $G=(V,E,f)$ be a pseudodigraph without isolated vertices, and let $n=\sum_{e\in{E}}f(e)$.
	Note that $|V|\leq{2|E|}\leq{2n}$.
	Then, the following algorithm can be successfully used to determine whether $G$ is connected.
	Specifically, it gets as input the set $E$ of edges, and returns ``\textit{YES}'' if and only if $G$ is
	connected, that is, $V$ cannot be divided into two disjoint sets $X,Y$ such that
	$\{(x,y) \mid x\in{X}$ and $y\in{Y}\}\cap{E}=\emptyset$.
\textup{
\\\\
Connected(set $E$)
\\
$\texttt{ 1.}$ \textbf{begin}
\\
$\texttt{ 2.}$ \hspace{20pt}$t:=0;$
\\
$\texttt{ 3.}$ \hspace{20pt}$sets:=();$
\\
$\texttt{ 4.}$ \hspace{20pt}\textbf{for each} $e\in{E}$ \textbf{do}
\\
$\texttt{ 5.}$ \hspace{20pt}\textbf{begin}
\\
$\texttt{ 6.}$ \hspace{40pt}$idx1:=0;$
\\
$\texttt{ 7.}$ \hspace{40pt}$idx2:=0;$
\\
$\texttt{ 8.}$ \hspace{40pt}\textbf{for} $i=1$ \textbf{to} $t$ \textbf{do}
\\
$\texttt{ 9.}$ \hspace{60pt}\textbf{for each} $x\in{sets.i}$ \textbf{do}
\\
$\texttt{10.}$ \hspace{60pt}\textbf{begin}
\\
$\texttt{11.}$ \hspace{80pt}\textbf{if} $x=e.1$ \textbf{then} $idx1:=i;$
\\
$\texttt{12.}$ \hspace{80pt}\textbf{if} $x=e.2$ \textbf{then} $idx2:=i;$
\\
$\texttt{13.}$ \hspace{60pt}\textbf{end}
\\
$\texttt{14.}$ \hspace{40pt}\textbf{if} $idx1=0$ then
\\
$\texttt{15.}$ \hspace{60pt}\textbf{if} $idx2=0$ then
\\
$\texttt{16.}$ \hspace{60pt}\textbf{begin}
\\
$\texttt{17.}$ \hspace{80pt}$sets:=sets\vartriangleleft{\{e.1,e.2\}};$
\\
$\texttt{18.}$ \hspace{80pt}$t:=t+1;$
\\
$\texttt{19.}$ \hspace{60pt}\textbf{end}
\\
$\texttt{20.}$ \hspace{60pt}\textbf{else} $sets.idx2:=sets.idx2\cup{\{e.1\}};$
\\
$\texttt{21.}$ \hspace{40pt}\textbf{else}
\\
$\texttt{22.}$ \hspace{60pt}\textbf{if} $idx2=0$ \textbf{then} $sets.idx1:=sets.idx1\cup{\{e.2\}};$
\\
$\texttt{23.}$ \hspace{60pt}\textbf{else}
\\
$\texttt{24.}$ \hspace{80pt}\textbf{if} $idx1\neq{idx2}$ \textbf{then}
\\
$\texttt{25.}$ \hspace{100pt}\textbf{if} $idx1<idx2$ \textbf{then}
\\
$\texttt{26.}$ \hspace{100pt}\textbf{begin}
\\
$\texttt{27.}$ \hspace{120pt}$M:=sets.idx1\cup{sets.idx2};$
\\
$\texttt{28.}$ \hspace{120pt}$sets:=sets\vartriangleright{idx2};$
\\
$\texttt{29.}$ \hspace{120pt}$sets:=sets\vartriangleright{idx1};$
\\
$\texttt{30.}$ \hspace{120pt}$sets:=sets\vartriangleleft{M};$
\\
$\texttt{31.}$ \hspace{120pt}$t:=t-1;$
\\
$\texttt{32.}$ \hspace{100pt}\textbf{end}
\\
$\texttt{33.}$ \hspace{100pt}\textbf{else}
\\
$\texttt{34.}$ \hspace{100pt}\textbf{begin}
\\
$\texttt{35.}$ \hspace{120pt}$M:=sets.idx1\cup{sets.idx2};$
\\
$\texttt{36.}$ \hspace{120pt}$sets:=sets\vartriangleright{idx1};$
\\
$\texttt{37.}$ \hspace{120pt}$sets:=sets\vartriangleright{idx2};$
\\
$\texttt{38.}$ \hspace{120pt}$sets:=sets\vartriangleleft{M};$
\\
$\texttt{39.}$ \hspace{120pt}$t:=t-1;$
\\
$\texttt{40.}$ \hspace{100pt}\textbf{end}
\\
$\texttt{41.}$ \hspace{20pt}\textbf{end}
\\
$\texttt{42.}$ \hspace{20pt}\textbf{if} $t=1$ \textbf{then return} $``\textit{YES}";$ 
\\
$\texttt{43.}$ \hspace{20pt}\textbf{else return} $``\textit{NO}";$
\\
$\texttt{44.}$ \textbf{end}
\\\\
}
	One can easily remark that we get an upper bound of $O(n^{2})$,
	and a lower bound of $\Omega(|E|)$.
\end{algorithm_}
\begin{algorithm_}
	Let $U$ be a $n$-uple whose components are strings of length two over a given alphabet.
	Then, the following algorithm returns the pseudodigraph associated to $U$ in linear time.
	Exactly, the time complexity of this algorithm is $\theta(n)$.
\textup{
\\\\
GetPseudodigraph(uple $U$, integer $n$)
\\
$\texttt{ 1.}$ \textbf{begin}
\\
$\texttt{ 2.}$ \hspace{20pt}$V:=\emptyset;$
\\
$\texttt{ 3.}$ \hspace{20pt}$E:=\emptyset;$
\\
$\texttt{ 4.}$ \hspace{20pt}\textbf{for} $i=1$ \textbf{to} $n$ \textbf{do}
\\
$\texttt{ 5.}$ \hspace{20pt}\textbf{begin}
\\
$\texttt{ 6.}$ \hspace{40pt}$V:=V\cup{\{\textit{First}(U.i,1),\textit{Last}(U.i,1)\}};$
\\
$\texttt{ 7.}$ \hspace{40pt}$E:=E\cup{\{(\textit{First}(U.i,1),\textit{Last}(U.i,1))\}};$
\\
$\texttt{ 8.}$ \hspace{20pt}\textbf{end}
\\
$\texttt{ 9.}$ \hspace{20pt}Let $f:E\rightarrow{\mathbb{N}}$ be such that $f(e)=0$ for all $e\in{E};$
\\
$\texttt{10.}$ \hspace{20pt}\textbf{for} $i=1$ \textbf{to} $n$ \textbf{do}
\\
$\texttt{11.}$ \hspace{40pt}$f(\textit{First}(U.i,1),\textit{Last}(U.i,1)):=f(\textit{First}(U.i,1),\textit{Last}(U.i,1))+1;$
\\
$\texttt{12.}$ \hspace{20pt}\textbf{return} $(V,E,f);$
\\
$\texttt{13.}$ \textbf{end}
}
\end{algorithm_}
\begin{algorithm_}
	Let $U$ be a $n$-uple as in the previous algorithm, and consider that
	$G(U)=(V,E,f)$.
	Note that $|V|\leq{2n}$. Then, the following algorithm returns ``YES'' if and only if exactly
	one of the following two conditions holds:
\begin{enumerate}
\item[1.] $F^{+}(v)=F^{-}(v)$ for all $v\in{V}$; 
\item[2.] There exist $2$ vertices $v_{1},v_{2}\in{V}$, $v_{1}\neq{v_{2}}$, such that
	$F^{+}(v_{1})=F^{-}(v_{1})+1$, $F^{-}(v_{2})=F^{+}(v_{2})+1$, and $F^{+}(v)=F^{-}(v)$
	for all $v\in{V-\{v_{1},v_{2}\}}$.
\end{enumerate}
\textup{
\\
TestConditions(pseudodigraph $(V,E,f)$)
\\
$\texttt{ 1.}$ \textbf{begin}
\\
$\texttt{ 2.}$ \hspace{20pt}\textbf{for each} $v\in{V}$ \textbf{do}
\\
$\texttt{ 3.}$ \hspace{20pt}\textbf{begin}
\\
$\texttt{ 4.}$ \hspace{40pt}$F^{+}(v):=0;$
\\
$\texttt{ 5.}$ \hspace{40pt}$F^{-}(v):=0;$
\\
$\texttt{ 6.}$ \hspace{20pt}\textbf{end}
\\
$\texttt{ 7.}$ \hspace{20pt}\textbf{for each} $e\in{E}$ \textbf{do}
\\
$\texttt{ 8.}$ \hspace{20pt}\textbf{begin}
\\
$\texttt{ 9.}$ \hspace{40pt}$F^{+}(e.1):=F^{+}(e.1)+f(e);$
\\
$\texttt{10.}$ \hspace{40pt}$F^{-}(e.2):=F^{-}(e.2)+f(e);$
\\
$\texttt{11.}$ \hspace{20pt}\textbf{end}
\\
$\texttt{12.}$ \hspace{20pt}$M:=\emptyset;$
\\
$\texttt{13.}$ \hspace{20pt}\textbf{for each} $v\in{V}$ \textbf{do}
\\
$\texttt{14.}$ \hspace{40pt}\textbf{if} $F^{+}(v)\neq{F^{-}(v)}$ \textbf{then} $M:=M\cup{\{v\}};$
\\
$\texttt{15.}$ \hspace{20pt}\textbf{if} $|M|=0$ \textbf{then return} $``\textit{YES}";$
\\
$\texttt{16.}$ \hspace{20pt}\textbf{else}
\\
$\texttt{17.}$ \hspace{40pt}\textbf{if} $|M|\neq{2}$ \textbf{then return} $``\textit{NO}";$
\\
$\texttt{18.}$ \hspace{40pt}\textbf{else}
\\
$\texttt{19.}$ \hspace{40pt}\textbf{begin}
\\
$\texttt{20.}$ \hspace{60pt}Let $M=\{v_{1},v_{2}\};$
\\
$\texttt{21.}$ \hspace{60pt}$d1:=F^{+}(v_{1})-F^{-}(v_{1});$
\\
$\texttt{22.}$ \hspace{60pt}$d2:=F^{+}(v_{2})-F^{-}(v_{2});$
\\
$\texttt{23.}$ \hspace{60pt}\textbf{if} $d1=1$ and $d2=-1$ \textbf{then return} $``\textit{YES}";$
\\
$\texttt{24.}$ \hspace{60pt}\textbf{else}
\\
$\texttt{25.}$ \hspace{80pt}\textbf{if} $d1=-1$ and $d2=1$ \textbf{then return} $``\textit{YES}";$
\\
$\texttt{26.}$ \hspace{80pt}\textbf{else return} $``\textit{NO}";$
\\
$\texttt{27.}$ \hspace{40pt}\textbf{end}
\\
$\texttt{28.}$ \textbf{end}
\\\\
}
	Note that the time complexity of the algorithm above is $\theta(|E|)$,
	and $|E|\leq{n}$, that is, the algorithm is linear.
\end{algorithm_}
\begin{algorithm_}
	We are now ready to describe the algorithm suggested by Theorem 5. Let $U$ be
	a $n$-uple whose components are strings of length two over a given alphabet.
	Then, the following polynomial algorithm returns ``\textit{YES}'' if and only if
	$U$ is a ``\textit{YES}'' instance of the \textup{\textsf{$(2,1)$--STRING--MATCH}} problem.
\textup{
\\\\
StringMatch$2$(uple $U$, integer $n$)
\\
$\texttt{ 1.}$ \textbf{begin}
\\
$\texttt{ 2.}$ \hspace{20pt}$G(U):=$GetPseudodigraph($U$,$n$)$;$
\\
$\texttt{ 3.}$ \hspace{20pt}\textbf{if} Connected($G(U).2$)$=``\textit{YES}"$ \textbf{then}
\\
$\texttt{ 4.}$ \hspace{40pt}\textbf{if} TestConditions($G(U)$)$=``\textit{YES}"$ \textbf{then return} $``\textit{YES}";$
\\
$\texttt{ 5.}$ \hspace{40pt}\textbf{else return} $``\textit{NO}";$
\\
$\texttt{ 6.}$ \hspace{20pt}\textbf{else return} $``\textit{NO}";$
\\
$\texttt{ 7.}$ \textbf{end}
\\\\
}
	It is easy to verify that we get an upper bound of $O(n^{2})$, and a lower bound
	of $\Omega(n)$. Thus, we conclude that the \textup{\textsf{$(2,1)$--STRING--MATCH}} problem
	can be solved in polynomial time.
\end{algorithm_}
\begin{example}
	Consider the $9$-uple $U=(ca,eb,ad,bf,dc,fe,ab,ab,ba)$. One can verify that
	$G(U)$ is connected and \textup{TestConditions($G(U)$)} returns ``\textit{YES}''. Thus, we conclude
	that $9$-uple $U$ is a ``\textit{YES}'' instance of the \textup{\textsf{$(2,1)$--STRING--MATCH}} problem,
	that is, there exists a path $P=(e_{1},\ldots,e_{9})$ in $G(U)$ such that
\begin{center}
	$|\{j \mid e_{j}=e\}|=|\{j \mid e=(\textit{First}(U.j,1),\textit{Last}(U.j,1))\}|$
\end{center}
	for all $e\in{G(U).2}$.
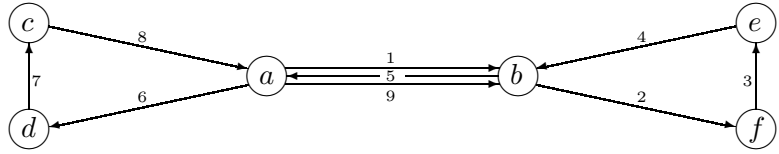
\begin{figure}[h]
\setlength{\unitlength}{1pt}
\begin{picture}(395,80)
	\put(120,40){\circle{15}} 
	\put(215,40){\circle{15}} 
	\put(30,60){\circle{15}}  
	\put(30,20){\circle{15}}  
	\put(305,60){\circle{15}} 
	\put(305,20){\circle{15}} 

	\put(117,37){$a$}
	\put(212,37){$b$}
	\put(27,57){$c$}
	\put(27,17){$d$}
	\put(302,57){$e$}
	\put(302,17){$f$}

	\put(207.5,40){\line(-1,0){35}}
	\put(162.5,40){\vector(-1,0){35}}
	\put(127,43.1){\vector(1,0){81}}
	\put(127,36.9){\vector(1,0){81}}
	\put(30,27.5){\vector(0,1){25}}
	\put(305,27.5){\vector(0,1){25}}
	\put(113,36.9){\rotatebox{12}{\vector(-1,0){77.2}}}
	\put(222,36.9){\rotatebox{348}{\vector(1,0){77}}}
	
	\put(37.3,59){\rotatebox{-12}{\vector(1,0){77.3}}}
	\put(297.2,59){\rotatebox{12}{\vector(-1,0){77.1}}}

	\put(165,45){\tiny{$1$}}
	\put(260,30){\tiny{$2$}}
	\put(260,53){\tiny{$4$}}
	\put(71,30){\tiny{$6$}}
	\put(71,53){\tiny{$8$}}
	\put(165,30.5){\tiny{$9$}}
	\put(300,36){\tiny{$3$}}
	\put(31,36){\tiny{$7$}}
	\put(165,38){\tiny{$5$}}
\end{picture}
\caption{$G(U)$: the pseudodigraph associated to $U$.}
\end{figure}

	For example, the path $P=(ab,bf,fe,eb,ba,ad,dc,ca,ab)$ illustrated above leads to the same conclusion.
\end{example}

\end{article}

\begin{thebibliography}{9}

\bibitem {cb:1}
Berge, C.
\newblock {\em Graphs}.
\newblock North-Holland, Amsterdam, 1985.

\bibitem {cb:2}
Berge, C.
\newblock {\em The Theory of Graphs}.
\newblock Dover, New York, 2001.

\bibitem {clrs:1}
Cormen, T. H., Leiserson, C. E., Rivest, R. L., C. Stein. 
\newblock {\em Introduction to Algorithms}.
\newblock The MIT Press, 2001.

\bibitem {rd:1}
Diestel, R.
\newblock {\em Graph Theory}.
\newblock Springer-Verlag, New York, 2000.

\bibitem {even:1}
Even, S.
\newblock {\em Graph Algorithms}.
\newblock Computer Science Press, 1979.

\bibitem {gy:1}
Gross, J. L., J. Yellen.
\newblock {\em Handbook of Graph Theory}.
\newblock CRC Press, 2003.

\bibitem {chp:1}
Papadimitriou, C. H.
{\em Computational Complexity\/}
(Addison-Wesley, 1993).

\end{thebibliography}
\end{document}